\documentclass[final,authoryear,5p,times,twocolumn]{elsarticle}

\usepackage{graphicx}
\usepackage{amssymb}
\usepackage{amsmath}
\usepackage{multirow}
\usepackage{color}
\journal{New Astronomy}
\begin{document}

\begin{frontmatter}

\title{Interactive Visualization of the Largest Radioastronomy Cubes}

\author{A.H. Hassan}\ead{ahassan@swin.edu.au}
\author{C.J. Fluke}\ead{cfluke@swin.edu.au}
\author{D.G. Barnes}\ead{dbarnes@swin.edu.au}

\address{Centre for Astrophysics \& Supercomputing, Swinburne University
of Technology, Hawthorn, Victoria, Australia}

\begin{abstract}

3D visualization is an important data analysis and knowledge discovery tool, however, interactive visualization of large 3D astronomical datasets poses a challenge for many existing data visualization packages. We present a solution to interactively visualize larger-than-memory 3D astronomical data cubes by utilizing a heterogeneous cluster of CPUs and GPUs. The system partitions the data volume into smaller sub-volumes that are distributed over the rendering workstations. A GPU-based ray casting volume rendering is performed to generate images for each sub-volume, which are composited to generate the whole volume output, and returned to the user. Datasets including the HI Parkes All Sky Survey (HIPASS - 12 GB) southern sky and the Galactic All Sky Survey (GASS - 26 GB) data cubes were used to demonstrate our framework's performance. The framework can render the GASS data cube with a maximum render time $< 0.3 $ second with $1024 \times 1024$ pixels output resolution using 3 rendering workstations and 8 GPUs. Our framework will scale to visualize larger datasets, even of Terabyte order, if proper hardware infrastructure is available.

\end{abstract}
\end{frontmatter}

% main text
\section{Introduction}

Visualization is the process of generating images of data in order to
aid knowledge discovery. Visualization is an integral part of
astronomy, playing a role in all stages of research (planning, data
monitoring, quality control, analysis, and interpretation) and
disemmination (publication, presentation, and public outreach).

Three-dimensional (3D) visualization has proven to be of great value
for studying and interpreting spectral data cubes from radiotelescopes \citep{norris:1994}, and more recently from optical telescopes fitted
with integral field units. A spectral data cube is a regular, scalar
3D data lattice. Two axes define (angular) sky
coordinates, and the third axis defines a (usually linear)
spectral abcissa. While sky coordinates can ordinarily be transformed
non-degenerately to a physical (spatial) representation, the same can
only be done for the spectral abcissa under special circumstances.
Nevertheless, it is standard practice to treat all three axes equally
and display both axially-aligned 2D slices, and arbitrary 3D
projections of the cube.

3D visualization of spectral data cubes can:
\begin{itemize}
\item give improved 3D perception of the data and enhanced
  comprehension of global properties (e.g.\ Figure~\ref{fig:HIPASS1}
  of this paper); 
\item be employed as a quality control tool to detect and investigate instrumental and data processing errors,
  \citep{oosterloo:1996,beeson:2004}; 
\item enable innovative quantitative data analysis through the
  selection and characterization of 3D regions and comparisons with
  simulations [e.g. \citet{fluke:2010}]; and 
\item support the discovery of strange phenomena, unexpected
  relations, or previously unidentified patterns that cannot be
  accomplished with automated techniques \citep{beeson:2004}.
\end{itemize}

\subsection{Volume rendering}

One particularly useful technique for studying 3D data volumes is {\em
  volume rendering}. Here, a color-coded 2D projection of the 3D data
is generated by emulating an optical model that describes the
interaction of light emitted, absorbed, or reflected by elements that
make up that volume [e.g. \citet{drebin:1988, lacroute:1994,
  levoy:1990, SCHWARZ:2007}]. Volume rendering gives the viewer a
global picture of a 3D dataset by displaying large and small-scale
features, as well as internal and external structures. While rendering
of isosurfaces or manually-segmented surfaces {\em can}\/ be used to
visualize spectral cubes, these techniques frequently fail for two
reasons: most astronomical sources do not have well-defined surfaces,
especially in the (typically) low signal-to-noise regime of spectral
line astronomy; and spectral cubes are not directly
representative of (or transformable to) a 3D spatial physical
representation, and so interpretation of the surface is difficult.  In
contrast to surface rendering, volume rendering remains useful where clear
feature segmentation cannot be done \citep{beeson:2003, gooch:1995,
  oosterloo:1995}.

One of the earliest volume rendering applications in astronomy was in
1992 by Domik and colleagues at the University of Colorado
\citep{domik:1992,Brugel:1993}. They introduced a preliminary
implementation, which they called ``translucent representation''.
Despite the limited graphics and processing capabilities available at
the time, they favored volume rendering over the other techniques
provided in their software suite (such as isosurfaces and data
slicing). Contemporary astronomy volume rendering implementations
include those that provide domain-specific transfer functions
[e.g. the hot gas shader \citep{oosterloo:1996}], those that offer effective
handling of adaptive grids and different data resolutions
\citep{kaehler:2006,nadeau:2001,magnor:2005}, and one that addresses
the {\em larger-than-memory}\/ data size problem \citep{beeson:2003}.

\subsection{The larger-than-memory problem}

The largest spectral line cubes from surveys carried out with
contemporary radiotelescopes are typically several gigabytes (GB) in
size.  For example, the image cube of the entire southern sky
generated from HIPASS data \citep[forthwith, the ``HIPASS cube'',][]{Barnes:2001} measures $1721 \times
1721 \times 1024$ voxels,\footnote{A {\em voxel}\/ is a volume element
  as a {\em pixel}\/ is a picture element.} and expressed as four-byte
floats occupies $\sim12$~GB in memory or on disk. The image cube
of the entire sky generated from GASS data \citep{Griffiths:2009}
measures $2502 \times 2501 \times 1093$ voxels, and occupies $\sim25$~GB.
The next generation of radiotelescopes [e.g.\ Australian Square
Kilometer Array Pathfinder (ASKAP)\footnote{http://www.atnf.csiro.au/SKA/}, LOw Frequency ARray (LOFAR)\footnote{http://www.lofar.org/}, Murchison Widefield Array (MWA)\footnote{http://www.mwatelescope.org/} , and Karoo Array Telescope (Meerkat)\footnote{http://www.ska.ac.za/meerkat/}] will produce even
larger cubes.  The standard, single-pointing spectral line cube from
ASKAP for example is expected to have dimensions of order $6144 \times 6144 \times 16384$,
occupying $\sim2.5$~terabytes (TB).

Volume rendering of these large cubes---from existing and planned
radiotelescopes---at interactive frame rates [i.e. better than $\sim5$
frames per second (fps)], is well beyond the capability of a single,
standalone workstation.  The principal limiting factor is memory
capacity, and so we refer to this problem as the {\em
  larger-than-memory}\/ data size problem.  One ``solution'' to the
larger-than-memory problem is to simply extract a sub-cube whose data
can fit in memory and accept that visualising the original spectral
cube in its entirety is not possible.  While there are circumstances
where this is an acceptable solution, it is sometimes impractical (a
cube may need to be visualized in 10 or more ``pieces''), and there
can be significant value in visualising a large data set in its
entirety.

\begin{figure*}
\begin{center}
\includegraphics[scale=0.8]{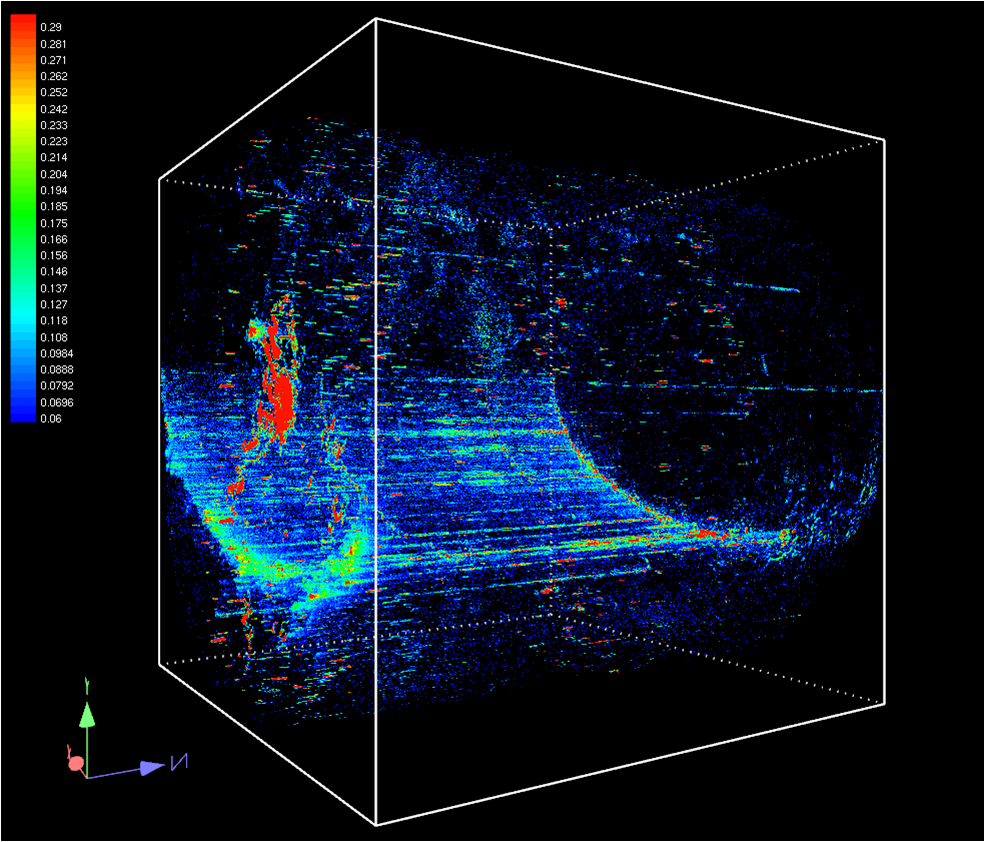}
\caption{Volume rendering of the HIPASS cube, accomplished with the
  approach described in this paper. The southern sky cube was generated by Russell Jurek (ATNF) from 387 individual cubes. See Section 1.2 for a description of the features in this cube.}
\label{fig:HIPASS1}                                 
\end{center}                                 
\end{figure*}

Consider for example the depiction of the HIPASS cube in
Figure~\ref{fig:HIPASS1}.  This volume rendering, accomplished with
the technique we present in this paper, provides a striking global
summary of not just the scientific content of the data---the
Magellanic Clouds and Stream (red feature near the centre of the
left-facing facet of the cube); the residual continuum emission, after
bandpass calibration, in the plane of the Milky Way Galaxy (the green
arc-like feature on the left-facing facet); and hundreds of galaxies
detected in the 21~cm neutral Hydrogen emission line (short bars
running along the spectral axis), which are weakly clustered and more
numerous nearby (towards the left)---but also the numerous artefacts
present in the processed data.  For example, residual continuum
emission from the Galaxy extends through the entire spectral space
(the ramp associated with the green arc-like feature), with an
intensity variation (``ripple'') along the spectral axis correlated on
the sky (e.g.\ the increased intensity on the upper part of the ramp
at the centre of the image).

The larger-than-memory problem has been previously examined in
astronomy by \citet{beeson:2003} who implemented the distributed
shear-warp algorithm \citep{lacroute:1994} over a Beowulf-style
cluster. In their solution ({\sc dvr}---distributed volume renderer),
the volume data was segmented and distributed to cluster nodes,
rendered locally, and the (sub-)images were combined by a controlling
node using an associative compositing operator.  {\sc Dvr}
demonstrated good rendering speeds compared to other solutions at the
time, but does not scale well to today's largest radioastronomy cubes
(see their Figure~9).  Even if {\sc dvr} scaled perfectly with no
parallelization costs, $\sim120$ 6-core Westmere processors\footnote{http://ark.intel.com/Product.aspx?id=47920\&processor=X5670\&spec-codes=SLBV7}, with each core handling 20~Mvox~sec$^{-1}$, would be needed to render the HIPASS
cube at 5~fps.\footnote{see \citet{beeson:2003} Equation~7.}  

\subsection{This work}

In this paper, we present a new solution to the larger-than-memory
problem, using a significantly smaller computer system than {\sc dvr}
requires for the same input image cube size.  Our objective is to provide
astronomers with a practical tool to interactively explore and
visualize the largest radioastronomy spectral data cubes in real time.
Our initial focus is visualizing data from ASKAP, however, it is also applicable to
facilities such as LOFAR, ALMA\footnote{http://www.alma.nrao.edu/}, MWA, and existing
large datasets such as the HIPASS cube. We begin with the HIPASS and GASS data cubes as benchmarks to test our solution performance and scalability. Both of these datasets are sufficiently large to provide a valid test of our volume rendering framework (see results section for more details).   

This work is about removing a potential technological barrier and enabling astronomers to have a qualitative look at their data as a first step to understanding the complex elements that will occur in the largest radioastronomy datacubes. In particular, we assert that global views can play a vital role as a quality control tool, especially since some of the upcoming facilities (e.g. ASKAP) will not be able to keep all the raw observational data after the initial processing phase. Being able to see the data products from such facilities in a real-time and interactive way may save a lot of precious time and data, and provide opportunities for reprocessing while the raw data products still exist.  Furthermore, we anticipate that global views may aid in discovering systematic and non-systematic noise effects, such as signals that vary across the sky due to calibration issues, which may otherwise take huge processing or data reduction effort to determine and extract (e.g. HIPASS data cube in figure \ref{fig:HIPASS1}). 

\section{Ray casting}

A volume rendering is generated by a process called {\em ray casting},
which computes the projection of a coloured, semi-transparent volume
onto a (finite) 2D viewing plane. The colour and the opacity of each
voxel are derived from its data value using a predefined mapping
operator called a ``transfer function".  For each pixel on the viewing
plane, the ray casting process computes a ray originating at this
pixel and projects it into the data volume. The ray is traced through
the volume, accumulating an aggregate colour and opacity which is
assigned to the pixel.  The process of compositing any two voxel
colours is also defined by the selected transfer function; see
\citet{levoy:1990} for a detailed description of the original ray
casting algorithm.  Although it is a computationally intensive task,
ray casting is trivially parallel.  This parallel nature has motivated
the development of number of parallel ray casting algorithms
[e.g. \citet{goel:1996,maximo:2008,scharsach:2005,Splotch:2010}].

Graphics Processing Units (GPUs) are the silicon processors used to
deliver the computations required for 3D computer graphics.  They
represent a cheap, commodity hardware (on graphics cards and
non-graphics co-processor cards) that can now be utilized as
massively-parallel, general purpose computational co-processors, by using software development kits such as CUDA\footnote{
  http://www.nvidia.com/object/ cuda\_what\_is.html} and
OpenCL\footnote{http://www.khronos.org/opencl/}.  GPUs have been taken up rapidly by the astronomy computing community [e.g. \citet{Hamada:2009,Wayth:2009,schive:2010}] and in the field of astrophysical
visualization, GPUs have been well-utilized for N-body particle
data \citep{kaehler:2006,li:2008,Szalay:2008,Splotch:2010,becciani:2010}.
In these examples, GPUs have been used to enhance the rendering speed
of the graphics primitives by e.g.\ $\sim23$\% for the Splotch code
\citep{Splotch:2010}, but these approaches are still limited to
datasets that fit within a single machine memory.

Motivated by the appearance of GPUs with large local memory (e.g.\
1.7~GB in the AMD ATI Radeon 5970; 1.5~GB in the NVIDIA GX285; up to
4~GB in the NVIDIA Tesla products), and the suitability of the
massively-parallel GPU architecture to the ray casting algorithm, we
developed a framework which utilizes a heterogeneous CPU and GPU
hardware infrastructure, combining shared- and distributed-memory
architectures, to yield a scalable volume rendering solution, capable
of volume rendering image cubes larger than a single machine memory
limit, in real-time and at interactive frame rates. We utilize GPUs as
massively-parallel floating point co-processors and consequently, our
framework may be suitable for other parallel scientific computing
applications, especially those processing or analysing
larger-than-memory images.

\subsection{Hardware and software architecture}
\begin{figure*}
\begin{center}
\includegraphics[scale=0.35]{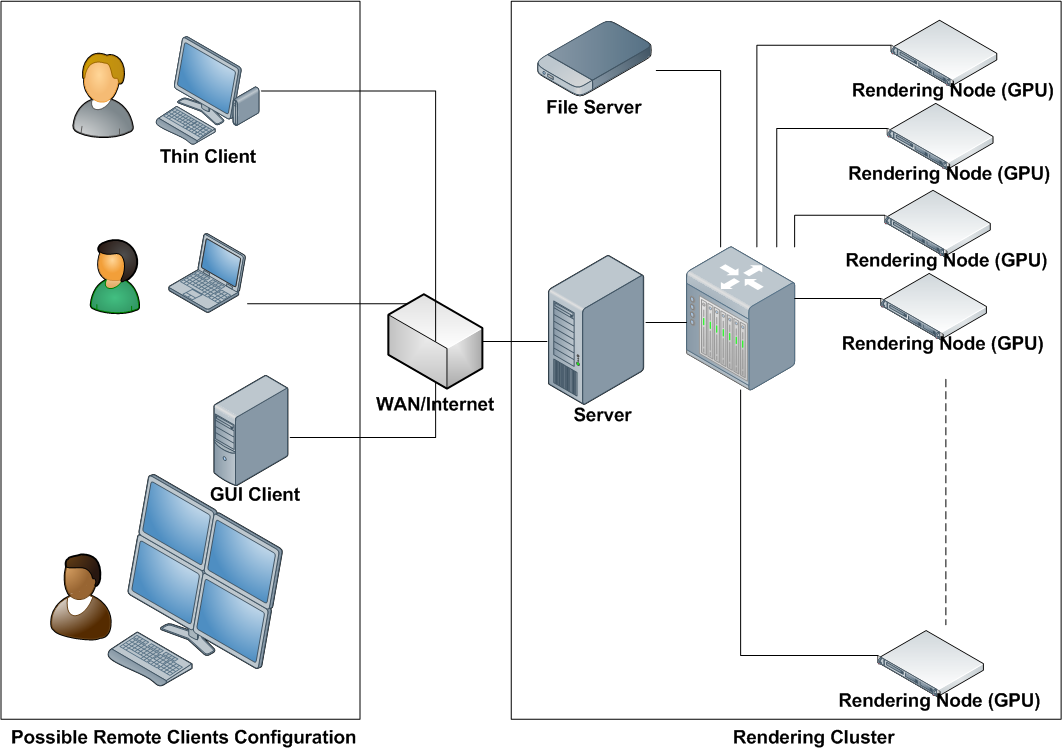}     
\caption{Conceptual diagram for the system hardware architecture. The main system layers are the viewer application, the rendering server, the data/file server, and the rendering client(s). The viewer application may work over different possible remote client's configuration including Web/Thin client, Desktop machine, or tiled display system.} 
\label{fig:conceptual} 
\end{center}      
\end{figure*}
To address the distributed volume rendering framework, we describe both the hardware and software architecture. Figure \ref{fig:conceptual} shows a conceptual diagram for our hardware architecture.  This architecture consists of the following main components:
\begin{enumerate}
\item  A Cluster of interconnected nodes, each featuring one or more GPU cards and one or more CPU cores;
\item  A Server machine, which will schedule tasks, exchange messages between the cluster and the viewer application, for final result composition;
\item  Viewer machine(s), which are a regular workstations associated with a display device and I/O mechanism with a connection to the server node; and
\item  A File Server, which is a storage node (can be physically any of the processing clients or the server machine) accessible by all the rendering nodes, the server, and the viewer machine.
\end{enumerate}
Based on this architecture configuration, we made the following design assumptions:
\begin{itemize}
\item  The viewer machine will have only a thin-client application with a small memory and processing requirements. Only a small percentage of the processing will be done on the viewer machine to facilitate the user interactivity, the I/O operations, and the result display. This will make the rendering cluster accessible over geographically distributed locations; 
\item  The final output at arbitrary  resolution, can be directed to a single display or a tiled-display system. Although this may decrease the final frame rate, the capability to view a dataset at its full resolution, or even with larger resolution, is very important. For example, some of the small features or details may be hidden if we view ASKAP-size data cubes ($6144 \times 6144$ spatial pixels are anticipated) with regular screen resolutions(e.g $1920 \times 1200$ for WUXGA is typical); and
\item  All the rendering nodes and the server can communicate using message passing interface (MPI) \footnote{ http://www.mcs.anl.gov/research/ projects/mpi/}. This can be achieved for either a static (the number of cluster nodes is constant) or dynamic cluster (the number of cluster nodes changes with the problem size).
\end{itemize}

The supporting software architecture utilizes the following software components:
\begin{itemize}
\item  MPI: for the communication between different rendering nodes and the rendering server;
\item  NVIDIA Compute Unified Device Architecture (CUDA): for the ray-casting implementation on the GPU;
\item  Multithreading and message queue: for the communication and management of different GPUs on the same rendering node; and
\item  TCP direct socket communication: to communicate between the viewer application and the server.
\end{itemize}

The components described above were integrated in a C++ framework to create a distributed GPU-based rendering farm. This C++ framework orchestrates the process of task distribution, data mapping, compositing, and communication between different GPU kernels, where the actual computation is done. This system assumes that the underlying hardware architecture is heterogeneous, which means that the number of GPUs in each rendering node, the amount of available memory, and possibly even processing power, is different from one node to the next. MPI is used for communication between the different rendering nodes and the rendering server. It has been used because it is the de-facto standard for distributed systems. Only the master thread (i.e. the thread responsible for the local scheduling and communication on each rendering node) on each processing node can communicate directly with the rendering server. Also, the server has no direct control over the processing threads (the threads associate to the GPU units on each node which handle the CUDA kernel invocation and result transfer between GPU memory and the node's system memory). 

The communication between the master thread and the processing threads on each of the rendering nodes is done using a shared Priority Message Queue in a completely asynchronous manner. This method speeds up the communication and minimizes the data sharing between the different threads. Also, it keeps the CUDA function calls within the same thread \footnote{sharing CUDA context between different thread is not supported as of version 2.3}. When required, any Master thread and the server can communicate together in a synchronous manner. The communication between the server and the viewer application is done through a direct TCP socket. This communication channel is opened and closed via the client and used to transfer control oriented messages and results. The current client implementation uses a QT\footnote{ http://qt.nokia.com/}-based user interface and OpenGL\footnote{ http://www.opengl.org/} for graphical display and user interaction. 

\subsection{Rendering overview}
The rendering process starts when the user selects a file to render (a menu item on the user interface). The client opens the requested file and loads the associated metadata. The file dimensions, and the recorded minimum and maximum data values are displayed to the user. The user may load the entire cube or a manually specified sub-cube. The user also has the control to use the minimum and the maximum value recorded in the file's metadata or to ask the server to recalculate the cube minimum and maximum.  Next, the file path and the selected cube dimensions are sent to the server to start the data loading process. The server performs a global scheduling task that partitions the cube according to the current rendering nodes. The server then sends a separate file-loading request to each of the processing clients. Each loading request contains the file path and the client's associated cube portion. In an asynchronous manner, the clients perform a local scheduling task that partitions its associated cube portion, depending on the available GPUs in each client, and start loading data. Once the data is loaded, the data is transferred from the system memory to the GPU memory. 

If requested by the user, each GPU calculates the minimum and the maximum of its cube portion, which are then sent back to the server where the global minimum and maximum is calculated.  The global minimum and maximum are returned to the client with a data loading confirmation message. The viewer application then generates the color map, which will be used in the rendering, and sends it back to the server as well. After these initialization steps are complete, the server will stay pending for render requests from the viewer application. Whenever the user interacts with the displayed output on the viewer application, the rendering parameters (the transformation and projection matrices) are sent to the server associated with a render request. This render request is distributed over the rendering nodes, which use the GPUs to generate the frame portions and combine (i.e. composite) them. The server performs the final frame composition and sends the result back to the client. The user  is also able to perform some other interactions, such as changing the color map, enabling/ disabling spectral channel(s), which are handled in the same manner.

\subsection{Data Partitioning}

\begin{figure*}
\begin{center}
\includegraphics[scale=0.3]{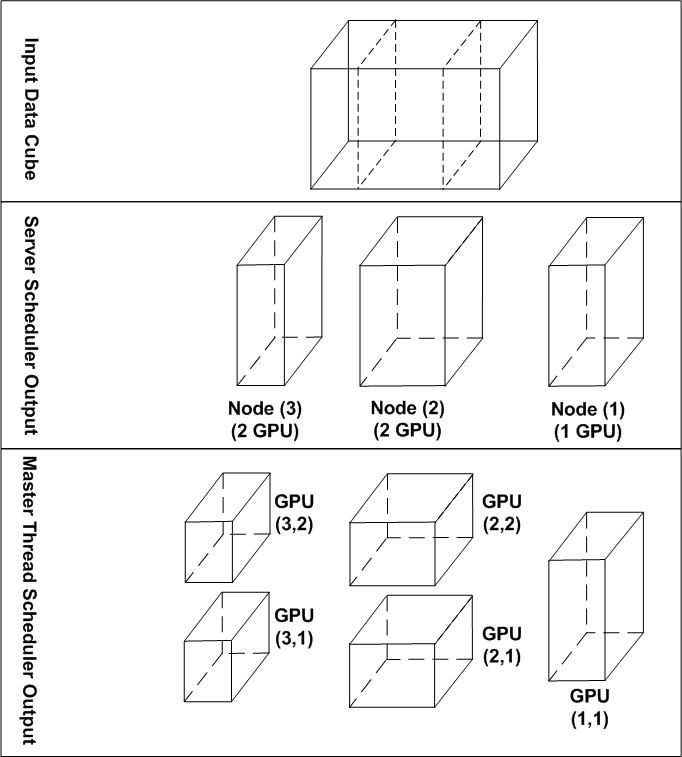}
\caption{Illustration of the data partitioning process over a cluster configuration of three processing clients. The first and the second processing client have two GPUs, while the third one contains one GPU. The cube partitioning is done based on the longest axis, so in this example the input data cube is partitioned into 3 parts over the X axis. For node 2 and node 3, the longest axis for their assigned cube is the Y axis. Node 1 has only one GPU so no further partition is required.}
\label{fig:datapartitioning}                                 
\end{center}                                 
\end{figure*}

The dataset is partitioned in an object-based manner, which means that each GPU gets only a portion of the data. At the same time each GPU is responsible for generating a portion of the final frame, corresponding to the projection of its assigned data on the current viewing plane using the ray-casting algorithm \citep{sabella:1988}. The final resultant frame images from each GPU are composited together to produce the final output frame. Figure \ref{fig:datapartitioning} shows an example data partitioning mechanism. Alternative data partitioning mechanisms could be applied, including more sophisticated techniques like binary space partitioning \citep{thibault:1987} or an Octree \citep{samet:1995}.  

The data partitioning mechanism affects the final rendering time and the overall performance as we show in Section 3. Based on the data partitioning schema and the viewing angle, the size of the rendering task differs from one GPU to another. Keeping the size of these tasks balanced is an important factor to achieve the highest possible frame rate and interactivity level.  We intend to leave further investigation and enhancement to the scheduler as  future work. In the current implementation, each scheduling module performs a separate data partitioning decision based on the current longest axis, as shown in Figure \ref{fig:datapartitioning}, so as to achieve a ``fair'' distribution. 

\subsection{Ray-Casting Process}

The processing threads are responsible for loading the data portion associated to its assigned GPU and transferring it to the GPU local memory. The GPU memory in our case is the most precious resource, because of the high time cost of data transfer between the GPU memory and the CPU memory, and the limitation of the GPU memory size (will reach 6 GB in the next Fermi GPU\footnote{ http://www.nvidia.com/object/ fermi\_architecture.html} , but is limited to 4 GB with the current Tesla cards). 

Each processing thread computes a ``rendering rectangle". This rendering rectangle represents the region that the associated GPU is responsible to fill with suitable colors in the final frame. This process is performed using the convex hull algorithm \citep{graham:1972}; the projection and transformation matrices; and previous knowledge about the extent of the data cube stored in the GPU's memory. Each ray within the rendering rectangle is tested against the bounding cube for the assigned data portion. In the case where the ray does not intersect with the cube, the current ray sampling process will exit. If the ray intersects with the cube, the entry and exit points are calculated and the sampling distance is determined based on the length of the ray segment, which will be inside the bounding cube. The ray casting kernel employs an early ray-termination mechanism \citep{levoy:1990} to speed up the overall rendering process. This mechamism terminates the ray casting process when the accumulated pixel value reach its maximum value.  The final output of the ray-casting process is a 2D floating point array of maximum recorded scalar value (in the case of maximum intensity projection\footnote{http://support.svi.nl/ wiki/MaximumIntensityProjection}) for each pixel. 

The usage of these optimization steps minimizes the number of threads needed and the amount of output buffer memory accessed by each GPU, thus speeding up the execution and memory transfer time. We note that these optimization steps do not affect the final output resolution or image quality. Some of these steps are also used to optimize the final frame composition steps, which we now describe.

\subsection{Image Composition}
\begin{figure*}
\begin{center}
\includegraphics[scale=0.35]{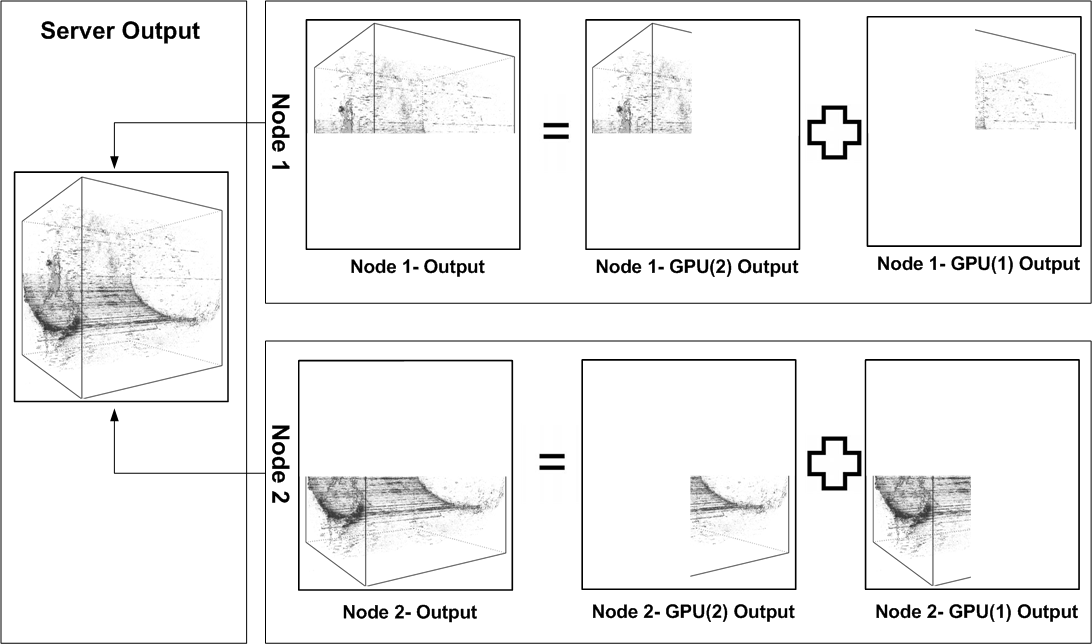}
\caption{Illustration of the image composition process over a cluster configuration of two processing clients each with two GPUs. A first compositing step occurs on each node, and the final image is combined on the server.}
\label{fig:imagecomposition}                                 
\end{center}                                 
\end{figure*}

The image composition is performed in two main stages (see Figure \ref{fig:imagecomposition}):
\begin{enumerate}
\item  Local stage at every rendering node. In this stage, each node is compositing the results generated by its GPUs to generate a single buffer. This composition is done with the guidance of the rendering rectangle for each GPU. Each GPU is executing a composition process only within its rendering rectangle on the final local frame buffer in a sequential manner; and
\item  Global stage at the server node. In this stage the server node is compositing the results of all the rendering nodes to generate the final rendering buffer.
\end{enumerate}

The use of this two stage process speeds-up the composition process and minimizes the processing effort required by the server and hence the final rendering time. The final composition complexity and validity depends on the selected transfer function. \citet{lombeyda:2001} provide a mathematical proof that the general alpha-blending volume rendering operator is associative, and can be applied in any blending order. 

\section{Results}

\begin{figure*}
\begin{center}
\includegraphics[scale=0.3]{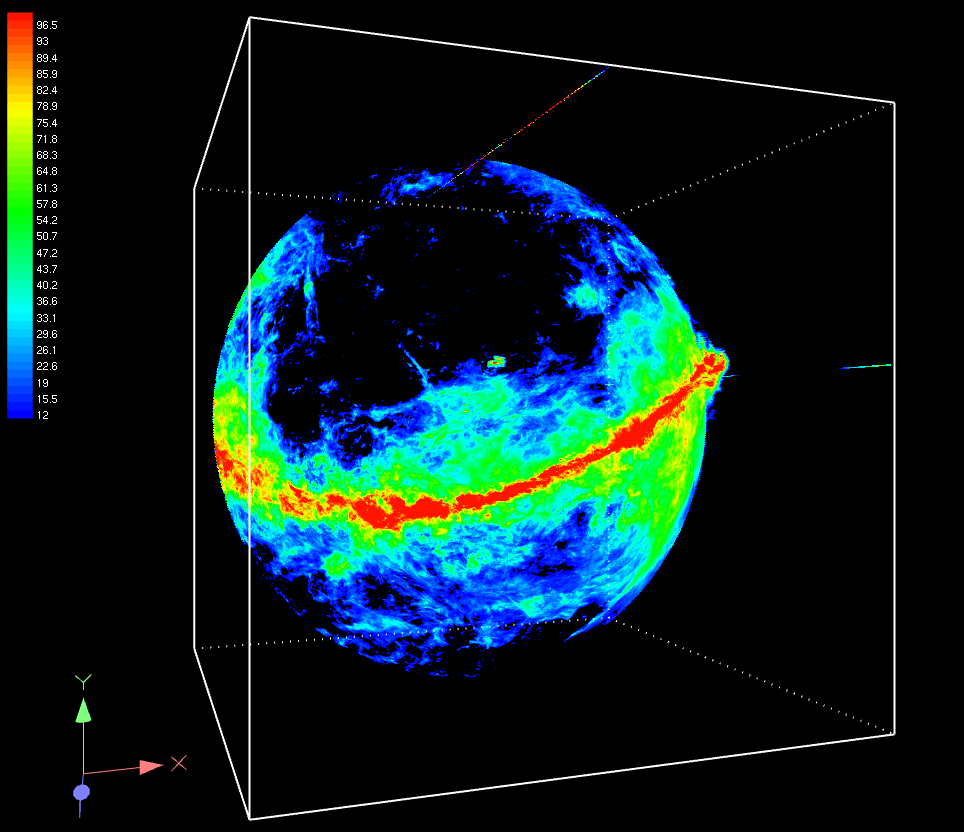}
\caption{Volume rendering of the GASS cube, accomplished with the approach described in this paper.}
\label{fig:GASS}                                 
\end{center}                                 
\end{figure*}
\begin{figure*}
\begin{center}
\includegraphics[scale=0.3]{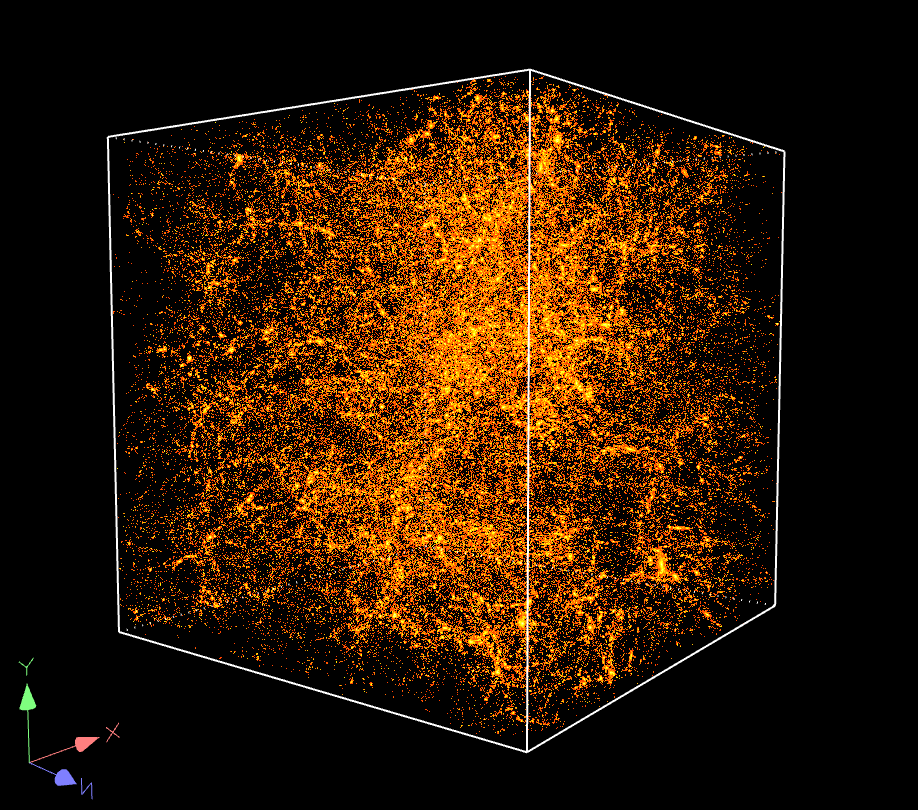}
\caption{Volume rendering of the Nbody cube, accomplished with the approach described in this paper. }
\label{fig:NBody}                                 
\end{center}                                 
\end{figure*}

A cluster of four interconnected workstations (the fourth always acting as a server) and nine GPUs (one associated with the server) was used to conduct framework timing tests. The hardware specifications of these workstations are shown in Table \ref{tab:ListOfPCs}. Details of the six individual  hardware configurations used are shown in Table \ref{tab:HWConfig}. Tests were performed using a $1024 \times 1024$ pixel viewport and a pre-computed sampling distance assuring that each voxel value (intersected by at least one ray) is sampled at least once. The communication is performed over a gigabit Ethernet network. Table \ref{tab:ListOfDatasets} shows the details of the datasets used for timing tests. Sample volume rendering are shown for HIPASS cube (Figure \ref{fig:HIPASS1}), GASS cube (Figure \ref{fig:GASS}), and a cube generated from a cosmological N-body simulation (Figure \ref{fig:NBody}, refer Table \ref{tab:ListOfDatasets}). Although the latter dataset is not from radio astronomy, it demonstrates the applicability of our framework to other 3D data volumes.   

We measure the total frame rendering time, $T_R$, which is the elapsed time between issuing a rendering request and receiving the rendered frame back. $T_R$ is an indication of the number of frames per second that the framework can render, and is a summation of the time spent on different framework sub-processes.

\begin{eqnarray}
\lefteqn{T_{R}= \rm{Max} (T_{\rm{ray}(i)}) + \rm{Max}(T_{\rm{merge}(j)})+ {}} \nonumber \\
& &  {}+ \rm{Max}(T_{\rm{comm}(j)}) + T_{\rm{server}} \pm \epsilon
\label{eq:totalrendering}
\end{eqnarray}
Where $1\leq i \leq \mbox{Number of GPUs}$, $1\leq j \leq \mbox{Number of Workstations}$,
$T_{\rm{ray}}$ indicates the time spent on the GPU doing  ray tracing, $T_{\rm{merge}}$ is the time spent on each client for local merging, $T_{\rm{comm}}$ is the communication time between each workstation and the server, and $T_{\rm{server}}$ is the time spent by the server doing the global merging. The last component, $\epsilon$, indicates that variation happens due to the overlapping effect and any unexpected delays (see below).

Due to the framework's distributed behaviour and the heterogeneous hardware, the maximum time spent in each sub-process is dominated by the time spent by the slowest processing element. For example, the ray casting process times vary between different GPUs based on the cube's orientation, which effects the size of the rendering rectangle of the dataset portion. This affects the number of rays that need to be traced and the path-length of the rays through the volume, which in turn affects the number of sampling operations required for each ray. Overlaps between communication and computation usually eliminate the differences between processing elements, but due to the randomness of this overlapping order, its effect is not constant.

\begin{figure*}
\begin{center}
\includegraphics[scale=0.5]{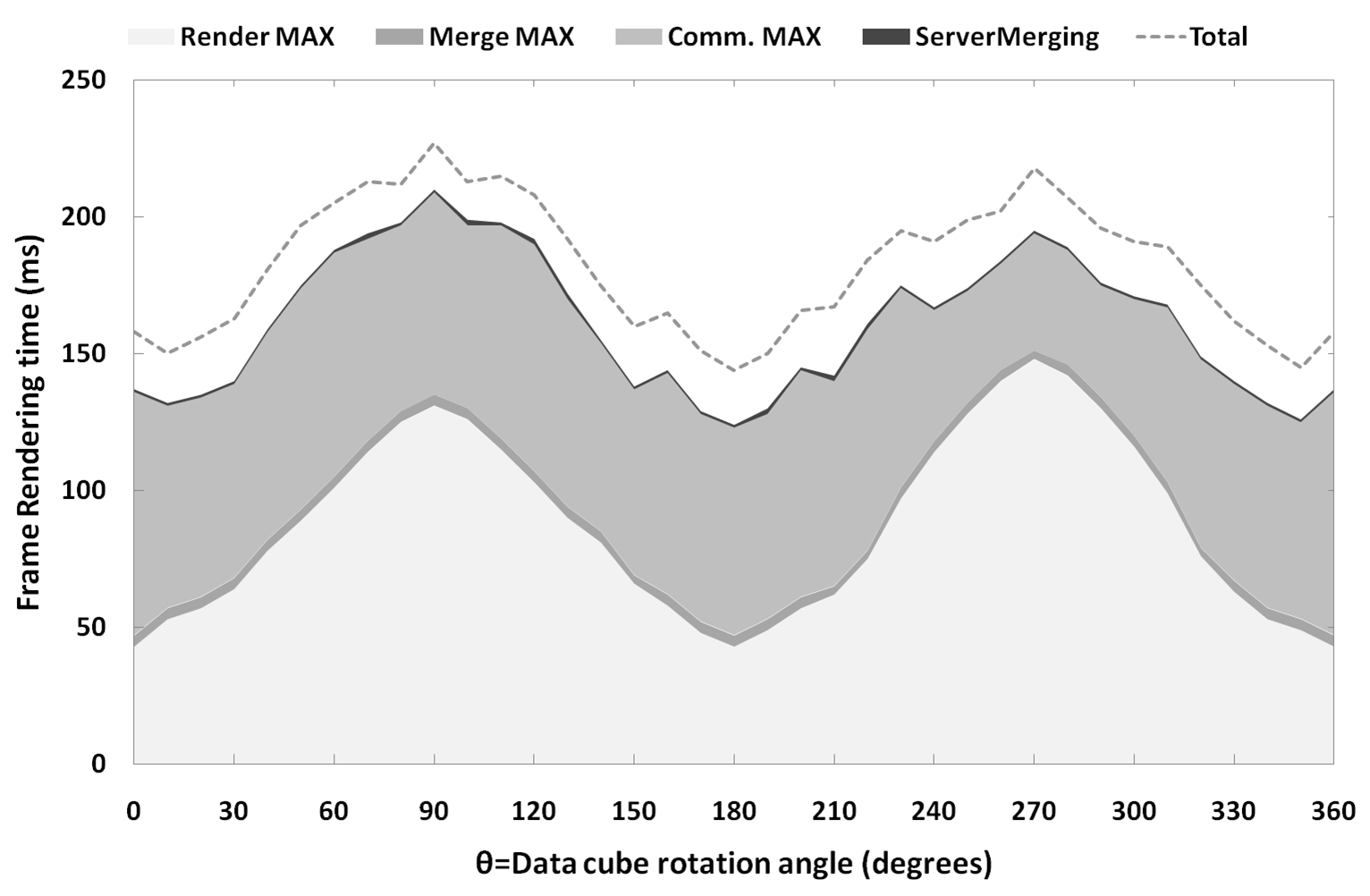}
\caption{Single frame rendering times for different cube rotation angles. The timing measurements were done for the HIPASS Southern Sky data cube on 3 workstations (one acting as a server) and 4 GPUs (2 P 4 G).}
\label{fig:HIPASSDetailedTiming}                                 
\end{center}                                 
\end{figure*}

Figure \ref{fig:HIPASSDetailedTiming} shows a sample timing diagram for the HIPASS cube as a function of cube orientation, using the 2P4G configuration (see Table \ref{tab:HWConfig}). The rendering time depends on the cube orientation, which is represented by the rotation angle, $\theta$, about the y-axis in degrees. $T_{ray}$ is the dominating factor for the variation in the rendering time with an average near 50\%. Also, a small variation between the frame rendering times for opposite angles, ($\theta$ and $180^\circ+\theta$), is caused by the early ray termination optimization step in the ray tracing implementation.   The value of $ \rm{Max}(\rm{T}_{\rm{merge}})$ , $ \rm{Max}(\rm{T}_{\rm{comm}})$, and $\rm{Max}(\rm{T}_{\rm{server}})$ are almost constant, because they are directly proportional to the final frame output size, which is not affected by the cube orientation.

\begin{figure*}
\begin{center}
\includegraphics[scale=0.5]{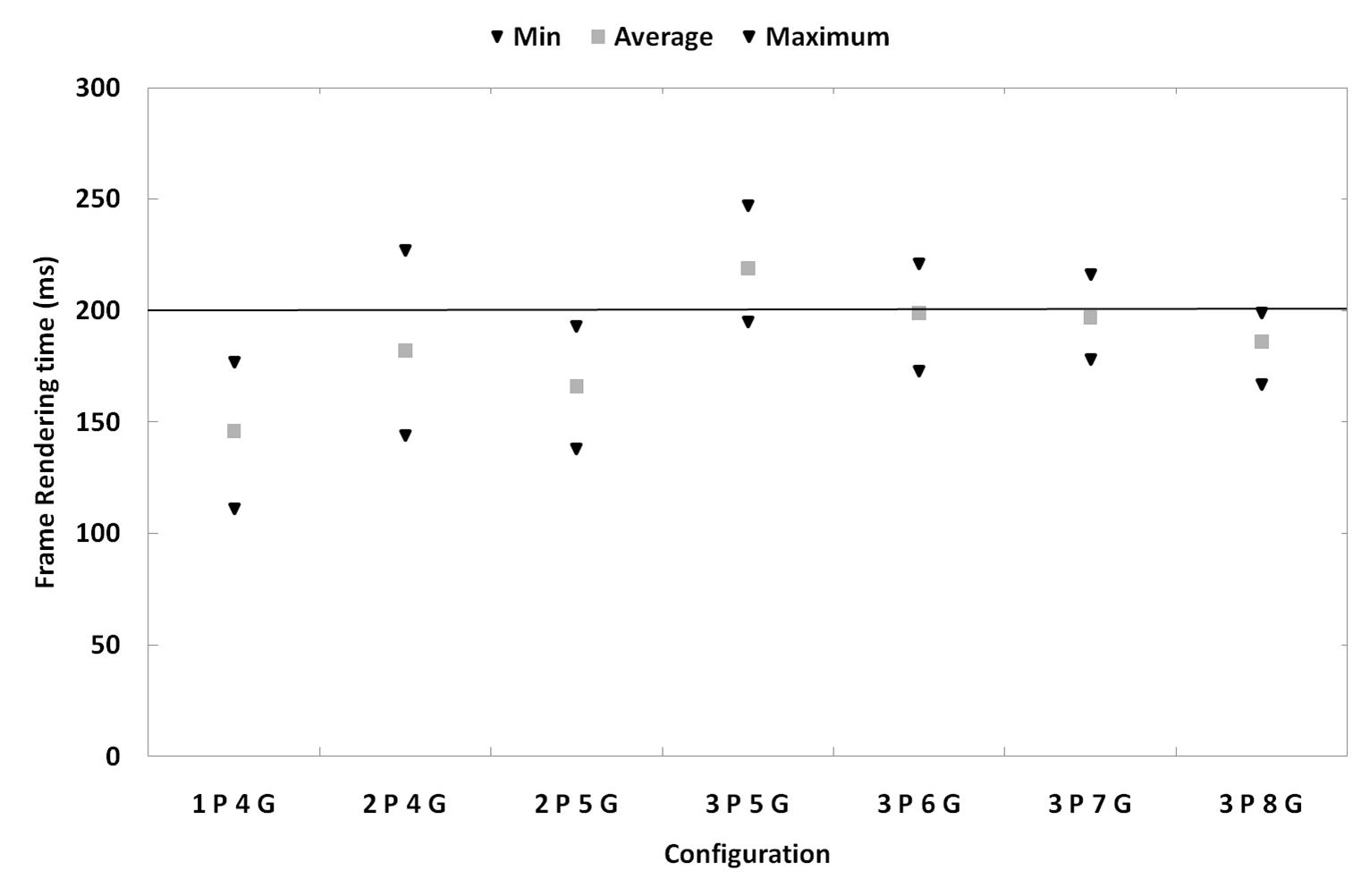}
\caption{Minimum, average, and maximum rendering time for the HIPASS data cube using different hardware configurations. The horizontal line shows our target of 5 fps for real-time interaction.}
\label{fig:HIPASSTiming}                                 
\end{center}                                 
\end{figure*}

\begin{figure*}
\begin{center}
\includegraphics[scale=0.5]{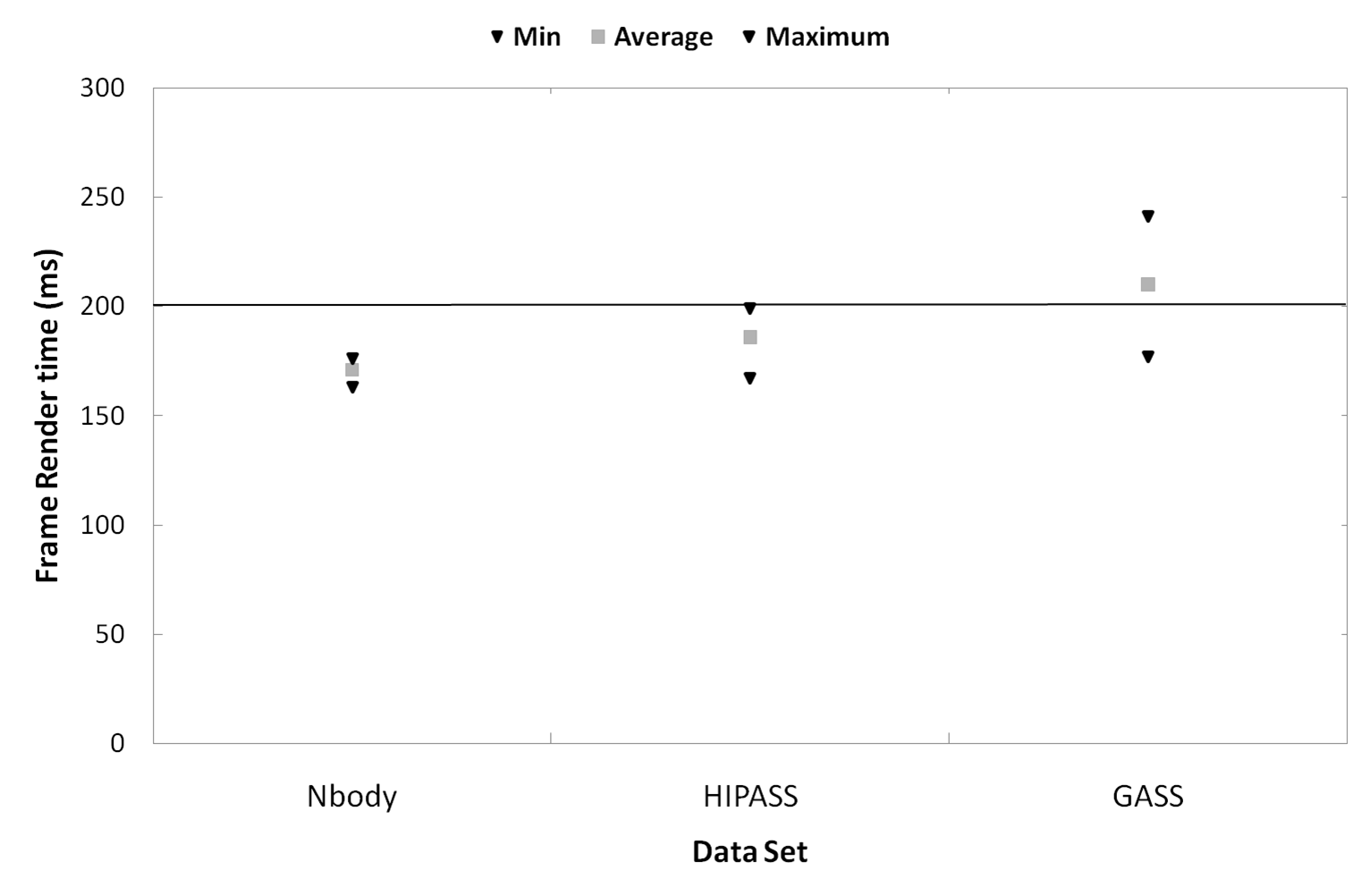}
\caption{Minimum, average, and maximum rendering time for the HIPASS cube, GASS Cube, and Nbody cube using the 3P8G configuration.}
\label{fig:AllTiming}                                 
\end{center}                                 
\end{figure*}

\begin{figure*}
\begin{center}
\includegraphics[scale=0.5]{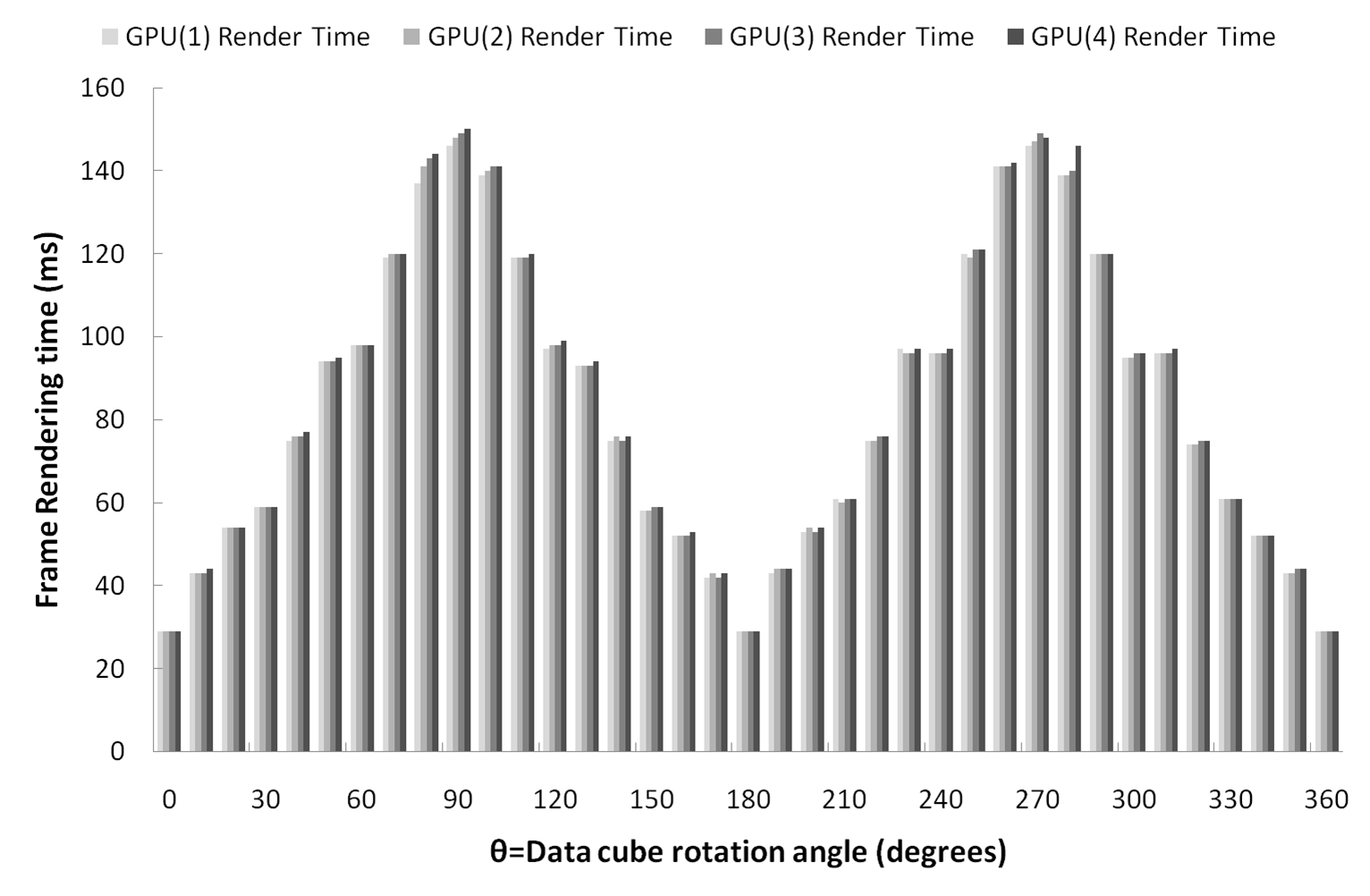}
\caption{Different GPU sub-frame rendering time for the HIPASS cube using the 2P2G configuration with early ray termination deactivated and orthographic projection.}
\label{fig:HIPASSOtho}                                 
\end{center}                                 
\end{figure*}

\begin{figure*}
\begin{center}
\includegraphics[scale=0.5]{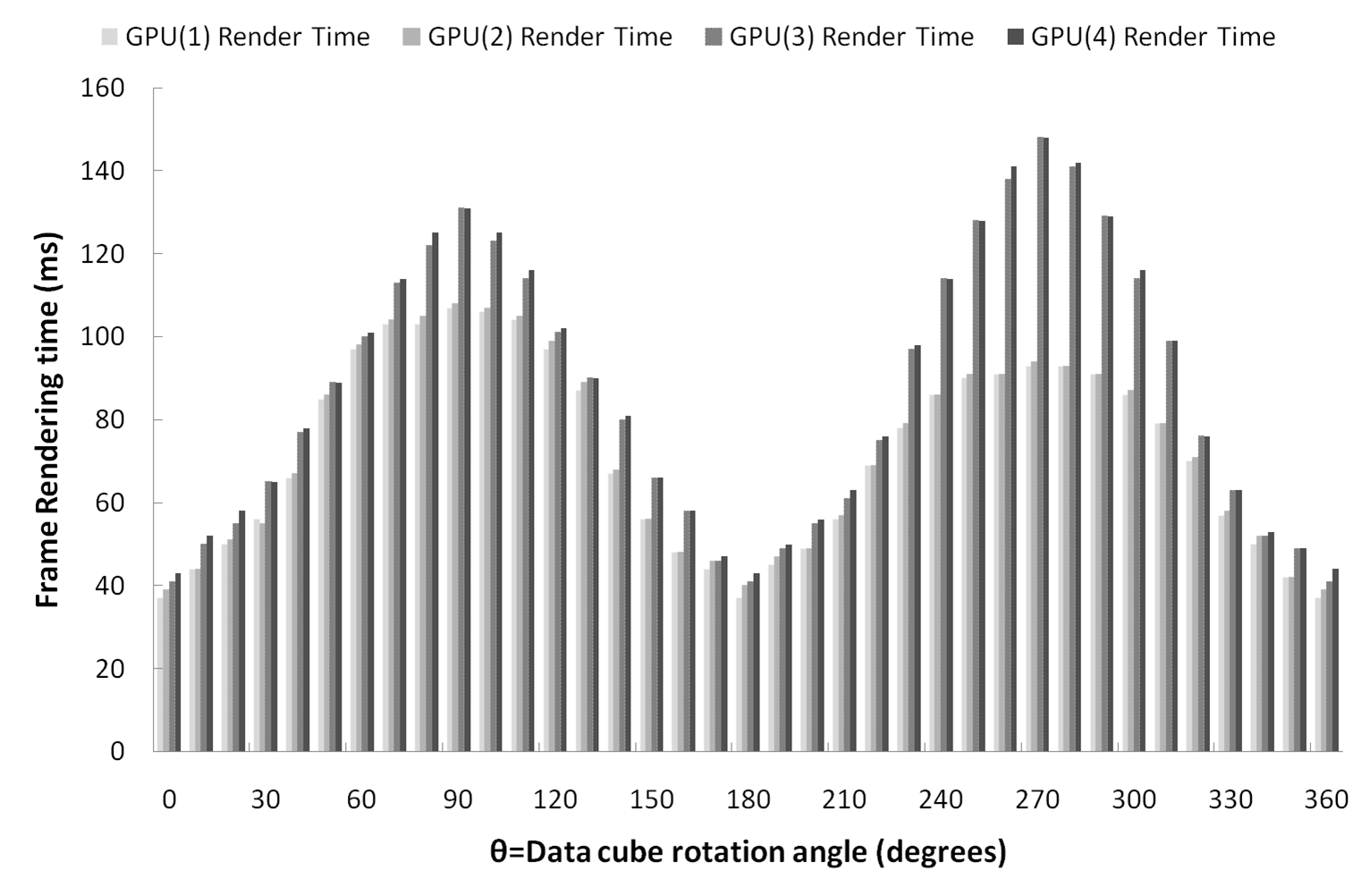}
\caption{Different GPU sub-frame rendering time for the HIPASS cube using the 2P2G configuration with early ray termination enabled and perspective projection.}
\label{fig:HIPASSPerspective}                                 
\end{center}                                 
\end{figure*}

Figure \ref{fig:HIPASSTiming} shows the average, minimum, and maximum frame rendering time for the HIPASS cube for different cluster configurations. Figure \ref{fig:AllTiming} shows the average, minimum, and maximum frame rendering time for the three datasets using the 3P8G configuration. Based on these timing tests, we can conclude that:
\begin{enumerate}
\item Increasing the number of GPUs does not necessarily reduce the final rendering time. Due to the communication and compositing overheads, for a fixed number of GPUs, the lower the number of workstations, the lower the frame rendering time.    

\item The unbalanced distribution of rendering tasks over the GPUs limits the framework speedup. Although different rendering tasks are being performed in parallel, the total frame rendering time is dominated by the maximum GPU rendering time as shown in equation 1. The current scheduler implementation uses the dataset dimensions and the GPU computational power (number of cores and memory size) to fairly partition the dataset over the rendering nodes and GPUs.  But the problem size each GPU tries to solve varies based on the dataset characteristics  (affects early ray termination), and the traced rays' length (because of perspective projection). Figures \ref{fig:HIPASSOtho} and \ref{fig:HIPASSPerspective} demonstrate this by showing the different sub-frame rendering time for each GPU for the HIPASS cube using the 2P4G configuration. Based on the variation in the difference between the maximum, the average, and the minimum rendering time, we expect this ``unbalanced distribution" influence to disappear with a large increase in the number of GPUs. On the other hand, the average total frame rendering time without early ray termination and with orthographic projection is higher by 12 \% from the average total frame rendering time with the early ray termination and perspective projection. The usage of a better data partitioning and scheduling which depends on the dataset characteristics and dimensions may also improve this behaviour. 

\end{enumerate}

\begin{figure*}
\begin{center}
\includegraphics[scale=0.5]{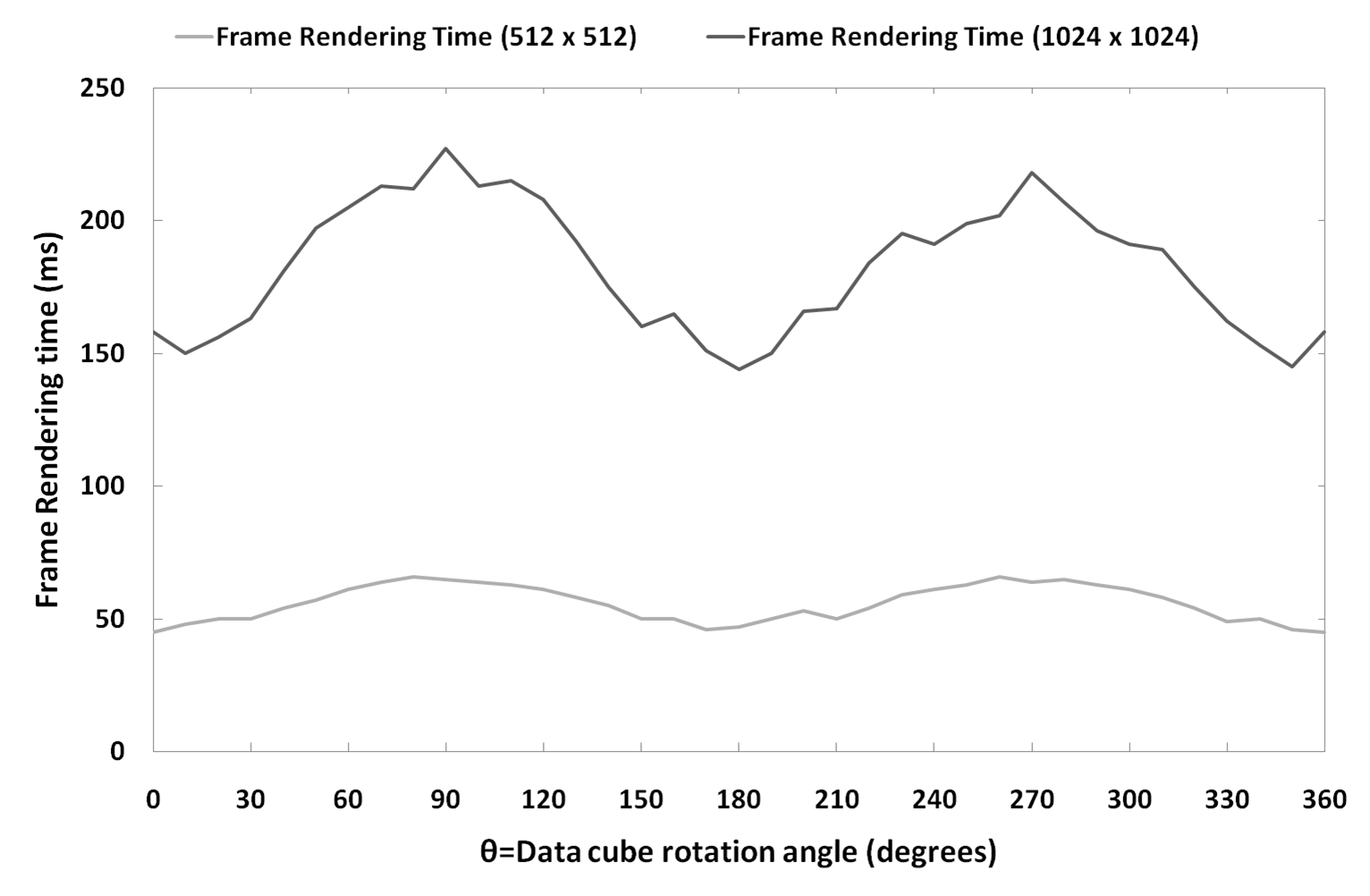}
\caption{Single frame rendering times for different cube rotation angles with different output sizes ( $1024 \times 1024$ and $512 \times 512$ pixels). The timing measurement was done for the HIPASS cube on 3 workstations (one acting as a server) and 4 GPUs (2 P 4 G).}
\label{fig:DifferentViewPort}                                 
\end{center}                                 
\end{figure*}

Figure \ref{fig:DifferentViewPort} shows timing for the HIPASS data cube with two different output sizes $1024 \times 1024$ and $512 \times 512$ pixels. The timing follows the same pattern for both output sizes but with an average time reduction of 69\%.  This reduction in the frame rendering time is due to the reduction in $T_{ray}$, the reduction in the size of merging operations, and the size of the data exchanged between the rendering workstations and the server.

\begin{table*}
\caption{Hardware specification for the cluster used to evaluate the performance of our framework.}
	\label{tab:ListOfPCs}
	\centering
		\begin{tabular}{ccccccc}
		\hline
			Index & GPU Model & GPU memory & Processor Model & System memory \\ \hline
			1 & 4x NVIDIA Tesla C1060&4 Gigabyte/GPU&	16x Intel Xeon X5550 &	18 Gigabyte\\ \hline
			2 & 2x NVIDIA Tesla C1060\footnote{http://www.nvidia.com/object/product\_tesla\_c1060\_us.html}&4 Gigabyte/GPU& 2x Nehalem i7 &	24 Gigabyte\\ \hline			3 &	4x NVIDIA Tesla C1060&4 Gigabyte/GPU&	2x Nehalem i7 &	24 Gigabyte\\ \hline		
			
			4 & 1x GeForce GTX 285& 1 Gigabyte&	1x Intel i7 930 &	12 Gigabyte\\ \hline		
		\end{tabular}

\end{table*}

\begin{table}
\caption{Details of the hardware configurations of the cluster (GPU+CPU) used to evaluate performance. The letter `P' is used as an abbreviation for the number of processing workstation, and the letter `G' is used as an abbreviation for the number of GPUs. The configurations do not include the server.}
	\label{tab:HWConfig}
	\centering
		\begin{tabular}{ccccc}
		\hline
			Configuration & Client(1) & Client(2) & Client(3) & Server \\ \hline
			1 P 4 G &	4 &  &  & 1\\ \hline				
			2 P 4 G &	2 & 2 &  & 1\\ \hline
			3 P 5 G &	2 & 2 & 1 & 1\\ \hline		
			3 P 6 G &	2 & 2 & 2 & 1\\ \hline
			3 P 7 G &	2 & 2 & 3 & 1\\ \hline
			3 P 8 G &	2 & 2 & 4 & 1\\ \hline	
		\end{tabular}	
\end{table}

\begin{table*}
\caption{Sample datasets used to evalute the performance of our framework.}
	\label{tab:ListOfDatasets}
	\centering
		\begin{tabular}{ccp{7cm}c}
		\hline
			Dataset Name & Dimensions (Data Points) & Source / Credits & File Size\\ \hline			
			Nbody cube  & 1024 x 1024 x 1024 & High resolution $1080^3$ dark matter simulation of a 125 Mpc/h box by Swinburne Computations for WiggleZ (SCWiggleZ) project (Poole et al 2010, in prep)  & 4 Gigabyte \\ \hline
			HIPASS Cube & 1721 x 1721  x 1025 & HIPASS Southern Sky, data courtesy Russell Jurek/HIPASS team & 12 Gigabyte \\ \hline
			GASS Cube & 2502 x 2501 x 1093 & The Parkes Galactic All-Sky Survey, data courtesy Naomi McClure-Griffiths/ GASS team \citep{Griffiths:2009} & 26 Gigabyte  \\ \hline
		\end{tabular}
	
\end{table*}

\section{Discussion}
\subsection{Framework performance and scalability}

In order to evaluate the performance gain from adopting GPUs as the main processing elements, we compared our performance timing with the distributed volume rendering implementation ({\sc dvr}) introduced by \citet{beeson:2003}.  The main performance evaluation done for {\sc dvr} used 2 GHz Pentium 4 CPUs and was able to render around 7 MVox/s. If we scale that to current CPU clock speeds (e.g. Westmere 2.93GHz processor), {\sc dvr} should be capable to render 120 MVox/s with 6 cores. Consequently, using equation (7) of \citet{beeson:2003} and by assuming perfect scalability, {\sc dvr} needs approximetely 120 $\times$ 6-core Westmere processors in order to render the HIPASS cube with 5 fps, while our framework can render that cube with 4 GPUs. 

During our performance and timing tests, we were unable to obtain a data file larger than the GASS data cube (26 GB). Furthermore, access to a larger GPU cluster was not available. Although we remain cautious about conclusions related to scalability of our framework, we believe it should be able to handle larger datasets, even of Terabyte order, if proper hardware infrastructure is available. We base this conclusion on the following:
\begin{itemize}
\item the amount of data transfer between the rendering workstation and the server is almost constant and depends on the output resolution rather than the input data size;  
\item the usage of ray casting combined with the rendering rectangle optimization makes the problem size almost constant for the rendering and the communication between the nodes; 
\item the overlapping between the computation and communication minimizes the communication delay; 
\item the two stage frame composition minimizes the amount of the processing required by the server and the amount of data exchanged between the clients and the server; and 
\item the local and global composition processes are done on GPU with a negligible cost compared to the rendering time.
\end{itemize}  

Moreover, the on-going increase in number of processing cores and size of local memory of GPUs  will help to decrease the rendering time and minimize the number of GPUs needed to render a certain data size. For example the next generation of NVIDIA Tesla GPUs is expected to have double the current number of processing elements and 1.5 times the current memory size.

\subsection{Future Work}

We expect the performance of our framework will be dramatically enhanced by using the next NVIDIA GPU architecture, code-named Fermi.  Features like predicated instructions, larger local memory, larger memory address space, greater DRAM bandwidth, improved instruction scheduler, and higher FLOP/S will provide our system with more powerful hardware infrastructure and remove some of the software bottlenecks.  Also, utilizing the direct integration between the Tesla cards and an InfiniBand network infrastructure may decrease the communication overhead and provide improved scalability\footnote{http://www.nvidia.com/object/ io\_1258539409179.html}.

To overcome the current GPU memory size limitation, we believe that lossy compression of the cube portions stored in the GPU local memory, using wavelet compression, will enable us to store and render larger cubes without affecting the final render quality. There have been a few trials to visualize datasets stored in the compressed wavelet format \citep{nguyen:2001,guthe:2002}, but to our knowledge none of these algorithms have been ported to GPU yet. It is likely that wavelet techniques can also be applied as a noise removal tool to increase the output quality. The usage of wavelet compression was demonstrated by \citet{pence:2009} with optical data,  however, some modifications are required to enable fast decompression and retrieval of random data positions \citep{nguyen:2001}. 

Quantitative data visualization support is still a missing ingredient from a complete visualization and analysis system for astronomy, and may be the main factor limiting a wider adoption of 3D visualization in astronomy. We aim to examine this further in future work.  A closely related issue is the use of noise-suppression techniques, as faint signals may be hidden in large-scale noise features. Designing one or more specialized transfer functions can provide the user with better visualization results. Most of the current well known transfer functions may not provide the user with the best visualization output because they were designed to serve other scientific domains. Transfer functions capable of suppressing noise and emphasising important data features will provide the users with better visualization outcomes and enhance the usefulness of 3D visualization as a data analysis and knowledge discovery tool.   

\section{Conclusion}
Visualization is a valuable tool for knowledge discovery. Along with providing insight and opportunities for analysis of sources under investigation, global views of data are vital for the detection of instrumentation errors, and the identification of data artefacts and noise characteristics.  New approaches are needed to visualize the massive, Terabyte order, data cubes that will be produced routinely by facilities such as ASKAP, MeerKat, LOFAR
and ultimately, the Square Kilometre Array.

In this work, we have introduced a framework to visualize larger than memory multispectral 3D datasets.  The framework provides the user with a real-time interactive volume rendering by combining between shared and distributed memory architectures, employing a distributed GPU infrastructure, and using the ray-casting volume rendering algorithm. We are trying to provide astronomers with a more affordable solution to deal with the upcoming data avalanche, by offering GPUs as an alternative to the currently used distributed computing infrastructures. By reducing the number of machines required to handle such datasets we not only reduce the overall hardware cost but also we provide an easier to deploy, and hence manage, solution. A remote viewer application is used to enable the user to control and interact with the framework. System implementation was done using QT, MPI, and CUDA within a C++ object oriented framework. 

Framework performance was evaluated using a cluster of four workstations and nine GPUs. The performance evaluation and timing tests were used to show the framework scalability, how different framework processes contribute to the final rendering time, and the effect of changing the cube orientation and the output viewport size on the rendering time. Medium size (12 GB) and relatively larger data cube (26 GB) were used throughout the timing tests. The maximum total rendering time for a 26 GB data cube with a $1024 \times 1024$ output viewport was $< 0.3$ second, with frame rates of 5 fps achievable. 

Based on the framework performance and timing analyses, we believe it should be able to visualize larger datasets, even of Terabyte order, if proper hardware infrastructure is available. The usage of ray casting, the overlapping between communication and computation, and the two stage results compositing minimize the parallelization overhead and the final frame rendering time.  
\section*{Acknowledgements}
We thank Dr. Virginia Kilborn, Dr. Emma Ryan-Weber, and Dr. Gregory Poole (Swinburne University of Technology), Dr. Russell Jurek and Dr. Naomi McClure-Griffiths  (ATNF - CSIRO), and Dr. Tara Murphy (Sydney University) for providing sample data cubes, useful discussions, and suggestions.

% \appendix

% \section{}
% \label{}

% Bibliographic references with the natbib package:
% Parenthetical: \citep{Bai92} produces (Bailyn 1992).
% Textual: \citet{Bai95} produces Bailyn et al. (1995).
% An affix and part of a reference:
%   \citep[e.g.][Ch. 2]{Bar76}
%   produces (e.g. Barnes et al. 1976, Ch. 2).
\section*{References}
\bibliographystyle{elsarticle-harv}        % Include this if you use bibtex 
\bibliography{references}           % and a bib file to produce the 
                                 % bibliography (preferred). The
                                 % correct style is generated by
                                 % Elsevier at the time of printing.

% \bibitem[Names(Year)]{label} or \bibitem[Names(Year)Long names]{label}.
% (\harvarditem{Name}{Year}{label} is also supported.)

%\bibitem[]{}
\end{document}